\documentclass{aa}
\usepackage{graphics,epsfig,astron,amsmath,amssymb,amstext}

\def\d {\mathrm{d}}
\def\Msun{\hbox{$M_{\odot}$}}
\def\tSW{\hbox{${t_{\rm SW}}$}}
\def\RSW{\hbox{${R_{\rm SW}}$}}
\def\tend{\hbox{${t_{\rm end}}$}}
\def\ESN{\hbox{${E_{\rm SN}}$}}
\def\Mej{\hbox{${M_{\rm ej}}$}}
\def\EdCR{\hbox{${\cal E}_{\rm CR}$}}

\def\pth{\hbox{${p_{\rm th}}$}}

\begin{document}

\thesaurus{12(02.01.1; 02.14.1; 09.19.2; 10.01.1)}

\title{Spallative Nucleosynthesis in Supernova Remnants}
  
\subtitle{I. Analytical Estimates}

\author{Etienne Parizot \and Luke Drury}

\institute{Dublin Institute for Advanced Studies, 5 Merrion Square, 
Dublin 2, Ireland\\ e-mail: parizot@cp.dias.ie; ld@cp.dias.ie}

\date{Recieved date; accepted date}

\maketitle

\begin{abstract}

Spallative nucleosynthesis is thought to be the only process capable 
of producing significant amount of Beryllium (Be) in the universe.  
Therefore, both energetic particles (EPs) and nuclei to be spalled 
(most efficiently C, N and O nuclei in this case) are required, which 
indicates that supernovae (SNe) may be directly involved in the 
synthesis of the Be nuclei observed in the halo stars of the Galaxy.  
We apply current knowledge relating to supernova remnant (SNR) 
evolution and particle shock acceleration to calculate the total Be 
yield associated with a SN explosion in the interstellar medium, 
focusing on the first stages of Galactic chemical evolution (i.e.  
when metallicity $0\le Z\le 10^{-2}$).  We show that dynamical aspects 
must be taken into account carefully, and present analytical 
calculations of the spallation reactions induced by the EPs 
accelerated at both the forward and the reverse shocks following the 
SN explosion.  Our results show that the production of Be in the early 
Galaxy is still poorly understood, and probably implies either 
selective acceleration processes (greatly favouring CNO acceleration), 
reconsideration of the observational data (notably the O~vs~Fe 
correlation), or even new energy sources.

\keywords{Acceleration of particles; Nuclear reactions, 
nucleosynthesis; ISM: supernova remnants; Galaxy: abundances}

\end{abstract}

\section{Introduction}

There has recently been considerable interest in the Galactic 
evolution of the abundances of the light elements Li, Be and B 
(Feltzing \& Gustafsson 1994; Reeves 1994; Cass\'e et al.  1995; 
Fields et al.  1994,1995; Ramaty et al.  1996,1997; Vangioni-Flam et 
al.  1998).  The Be abundance is particularly interesting because this 
element is thought to be produced exclusively by spallation reactions 
involving collisions between nuclei of the CNO group of elements and 
protons or alpha particles at energies greater than about $30\,\rm 
MeV$ per nucleon (MeV/n).  Thus the evolution of the Be abundance 
contains information about the particle acceleration and cosmic ray 
history of the Galaxy.

The evolution of the Be abundance, and indeed the evolution of all 
elemental abundances, has to be deduced from observations of the 
fossil abundances preserved in the oldest halo stars.  Advances in 
spectroscopy over the last decade have greatly improved the quality of 
the data available (Duncan et al.  1992,1997; Edvardsson et al.  1994; 
Gilmore et al.  1992; Kiselman \& Carlsson 1996; Molaro et al.  1997; 
Ryan et al.  1994) and the main result is easily summarized~: in old 
halo stars of low metallicity, the ratio of the Be abundance to the 
Iron (Fe) abundance appears constant, that is to say the Be abundance 
rises {\em linearly} with the Fe abundance.

This has been a surprising result.  Naively one had expected that, 
because Be is a secondary product produced from the primary CNO 
nuclei, its abundance should vary {\em quadratically} as a function of 
the primary abundances at low metallicities.  Indeed, considering that 
the cosmic rays (CRs) responsible for the Be production are somehow 
related to the explosion of supernovae (SNe) in the Galaxy, it is 
natural to assume that their flux is proportional to the SN rate, $\d 
N_{\mathrm{SN}}/\d t$.  Now since the number of CNO nuclei present in 
the Galaxy at time $t$ is proportional to the total number of SN 
having already exploded, $N_{\mathrm{SN}}(t)$, the Be production rate 
has to be proportional to $N_{\mathrm{SN}}\d N_{\mathrm{SN}}/\d t$.  
Therefore, the integrated amount of Be grows as $N_{\mathrm{SN}}^{2}$, 
that is quadratically with respect to the ambient metallicity (C,N,O 
or Fe, assumed to be more or less proportional to one another).

The above reasoning, however, relies on two basic assumptions that 
need not be fulfilled~: i) the CRs recently accelerated interact with 
all the CNO nuclei already produced and dispersed in the entire Galaxy 
and ii) the CRs are made of the ambient material, dominated by H and 
He nuclei.  Instead, it might be i) that the proton rich CRs recently 
accelerated interact predominantly with the freshly synthesized CNO 
nuclei near the explosion site and ii) that a significant fraction of 
the CNO rich SN ejecta are also accelerated.  In both cases, a linear 
growth of the Be abundance with respect to Fe or O would arise very 
naturally, since the number of Be-producing spallation reactions 
induced by each individual supernova would be directly linked to its 
local, individual CNO supply, independently of the accumulated amount 
of CNO in the Galaxy.

In fact, as emphasised by Ramaty et al.  (1997), the simplest 
explanation of the observational data is to assume that each 
core-collapse supernova produces on average 0.1~\Msun~of Fe, one to 
few \Msun~of the CNO elements and $2.8\times10^{-8}\,\Msun$ of Be, 
{\em with no metallicity dependence}.  Clearly if this is the case and 
the production of Be is directly linked to that of the main primary 
elements, the observed linear relation between Be and Fe will be 
reproduced whatever the complications of infall, mixing and outflow 
required by the Galactic evolution models.  On the other hand, 
although the simplest explanation of the data is clearly to suppose a 
primary behaviour for the Be production, it is possible that this 
could be an artifact of the evolutionary models (as argued, e.g., by 
Casuso \& Beckman 1997).

Most work in this area has attempted to deduce information about 
cosmic ray (or other accelerated particle populations) in the early 
galaxy by working backwards from the abundance observations.  While 
perfectly legitimate, our feeling is that the observational errors and 
the uncertainties relating to Galactic evolution in general make this 
a very difficult task.  We have chosen to approach the problem from 
the other direction and ask what currently favoured models for 
particle acceleration in supernova remnants (SNRs) imply for light 
element production.  This is in the general spirit of recent 
calculations of the $\pi^0$-decay gamma-ray luminosity of SNRs (Drury 
et~al.  1994) and the detailed chemical composition of SNR shock 
accelerated particles (Ellison, Drury \& Meyer, 1997) where we look 
for potentially observable consequences of theoretical models for 
cosmic ray production in SNRs.

Interestingly enough, the study of particle acceleration in SNRs 
suggests that both alternatives to the naive scenario mentioned above 
do occur in practice, as demonstrated qualitatively in 
Sect.~\ref{PartAccInSNRs}.  The first of these alternatives, namely 
the local interaction of newly accelerated cosmic rays in the vicinity 
of SN explosion sites, has already been called upon by Feltzing \& 
Gustaffson (1994), as well as the second, the acceleration of enriched 
ejecta through a SN reverse shock, by Ramaty et al.  (1997).  However, 
no careful calculations have yet been done, taking the dynamics of the 
process into account, notably the dilution of SN ejecta and the 
adiabatic losses.  Yet we show below that they have a crucial 
influence on the total amount of Be produced, and that a 
time-dependent treatment is required.  Indeed, the evolution of a SNR 
is essentially a dynamical problem in which the acceleration rate as 
well as the chemical composition inside the remnant are functions of 
time.  The results of the full calculation of both processes and the 
discussion of their implications for the chemical evolution of the 
Galaxy will be found in an associated paper (Parizot \&~Drury 1999).  
Here we present simple analytical calculations which provide an 
accurate understanding of the dynamics of light element production in 
SNRs and elucidates the role and influence of the different 
parameters, notably the ambient density.

Although Li and B are also produced in the processes under study, we 
shall choose here Be as our `typical' light element, because nuclear 
spallation of CNO is thought to be its only production mechanism, 
while Li is also (and actually mainly) produced through $\alpha + 
\alpha$ reactions, $^{7}$Li may be produced partly in AGB stars (Abia 
et~al.  1993), and $^{11}$B neutrino spallation may be important as 
seems to be required by chemical evolution analysis (Vangioni-Flam 
et~al.  1996).  In order to compare our results with the observations, 
we simply note that, as emphasized in Ramaty et al.  (1997), the data 
relating to the Galactic Be evolution as a function of [Fe/H] indicate 
that $\sim 1.6\,10^{-6}$ nuclei of Be must be produced in the early 
Galaxy for each Fe nucleus.  Therefore, if Be production is indeed 
induced, directly or indirectly, by SNe explosions, and since the 
average SN yield in Fe is thought to be $\sim 0.11\,\Msun$, each 
supernova must lead to an average production of $\sim 3.8\,10^{48}$ 
nuclei (or $\sim 2.8\,10^{-8}\,\Msun$) of Be, with an uncertainty of 
about a factor of~2 (Ramaty et al.  1997).  We adopt this value as the 
`standard needed number' of Be per supernova explosion.  To state this 
again in a different way, for an average SN yield in CNO of, say, 
$\sim 1\,\Msun$, the required spallation rate per CNO atom is $\sim 
3\,10^{-8}$.

\section{Particle acceleration in SNRs}
\label{PartAccInSNRs}

It is generally believed that cosmic ray production in SNRs occurs
through the process of diffusive shock acceleration operating at the
strong shock waves generated by the interaction between the ejecta
from the supernova explosion and the surrounding medium. Significant
effort has been put into developing dynamical models of SNR evolution
which incorporate, at varying levels of detail, this basic
acceleration and injection process (one of the major advantages of
shock acceleration is that it does not require a separate injection
process). Qualitatively the main features can be crudely summarised as
follows.

In a core collapse SN the collapse releases roughly the gravitational
binding energy of a neutron star, some $10^{53}\,\rm erg$, but most of
this is radiated away in neutrinos. About $\ESN = 10^{51}\,\rm erg$ is
transferred, by processes which are still somewhat obscure, to the
outer layers of the progenitor star which are then ejected at
velocities of a few percent of the speed of light. Initially the
explosion energy is almost entirely in the form of kinetic energy of
these fast-moving ejecta. As the ejecta interact with the surrounding
circumstellar and interstellar material they drive a strong shock
ahead into the surrounding medium. The region of very hot high
pressure shocked material behind this forward shock also drives a
weaker shock backwards into the ejecta giving rise to a characteristic
forward reverse shock pair separated by a rather unstable contact
discontinuity.

This initial phase of the remnant evolution lasts until the amount of
ambient matter swept up by the remnant is roughly equal to the
original ejecta mass. At this so-called sweep-up time, \tSW, the
energy flux through the shocks is at its highest, the expansion of the
remnant begins to slow down, and a significant part of the explosion
energy has been converted from kinetic energy associated with the bulk
expansion to thermal (and non-thermal) energy associated with 
microscopic degrees of freedom of the system. The remnant now enters
the second, and main, phase of its evolution in which there is rough
equipartition between the microscopic and macroscopic energy
densities. The evolution in this phase is approximately self-similar
and resembles the exact solution obtained by Sedov for a strong point
explosion in a cold gas. 

It is important to realise that the approximate equality of the energy
associated with the macroscopic and microscopic degrees of freedom in
the Sedov-like phase is not a static equilibrium but is generated
dynamically by two competing processes. As long as the remnant is
compact the energy density, and thus pressure, of the microscopic
degrees of freedom is very much greater than that of the external
medium. This strong pressure gradient drives an expansion of the
remnant which adiabatically reduces the microscopic degrees of freedom
of the medium inside the remnant and converts the energy back into
bulk kinetic energy of expansion. At the same time the strong shock
which marks the boundary of the remnant converts this macroscopic
kinetic energy of expansion back into microscopic internal form. Thus
there is a continuous recycling of the original explosion energy
between the micro and macro scales. This continues until either the
external pressure is no longer negligible compared to the internal, or
the time-scales become so long that radiative cooling becomes
important. The time scales for the conversion of kinetic energy to
internal energy and vice versa are roughly equal and of order the
dynamical time scale of the remnant which is of order the age of the
remnant, hence the approximately self-similar evolution. 

In terms of particle acceleration the theory assumes that strong 
collisionless shocks in a tenuous plasma automatically and 
inevitably generate an approximately power law distribution of 
accelerated particles which connects smoothly to the shock-heated 
particle distribution at `the\-rmal' energies and extends up to a 
maximum energy constrained by the shock size, speed, age and magnetic 
field.  The acceleration mechanism is a variant of Fermi acceleration 
based on scattering from magnetic field structures on both sides of 
the shock.  A key point is that these scattering structures are not 
those responsible for general scattering on the ISM, but strongly 
amplified local structures generated in a non-linear bootstrap process 
by the accelerated particles themselves.  As long as the shock is 
strong it will be associated with strong magnetic turbulence which 
drives the effective local diffusion coefficient down to values close 
to the Bohm value.  As pointed out by Achterberg et al.  (1994)  the 
extreme sharpness of the radio rims of some shell type SNRs can be 
interpreted as observational evidence for this type of effect.  The 
source of free energy for the wave excitation is of course the strong 
gradient in the energetic particle distribution at the edge.  Thus in 
the interior of the remnant, where the gradients are absent or much 
weaker, we do not expect such low values of the diffusion coefficient.

The net effect is that the edge of the remnant, as far as the
accelerated particles are concerned, is both a self-generated diffusion
barrier and a source of freshly accelerated particles. Except at the
very highest energies the particles produced at the shock are
convected with the post-shock flow into the interior of the remnant
and effectively trapped there until the shock weakens to the point
where the self-generated wave field around the shock can no longer be
sustained. At this point the diffusion barrier collapses and the
trapped particle population is free to diffuse out into the general
ISM.

In terms of bulk energetics, the total energy of the accelerated 
particle population is low during the first ballistic phase of the 
expansion (because little of the explosion energy has been processed 
through the shocks) but rises rapidly as $t\approx\tSW$.  During the 
sedov-like phase the total energy in accelerated particles is roughly 
constant at a significant fraction of the explosion energy (0.1 to 0.5 
typically).  However, this is because of the dynamic recycling 
described above.  Any individual particle is subject to adiabatic 
losses on the dynamical time-scale of the remnant, while the 
energy lost this way goes into driving the shock and thus generating 
new particles, distributed over the whole energy spectrum.

\section{Spallation reactions within SNRs}

\subsection{Qualitative overview}

We now turn to the production of Li, Be and B (LiBeB) by spallation 
reactions within a SNR. As emphasized above, there are two obvious 
mechanisms.  One is the irradiation of the CNO ejecta by accelerated 
protons and alphas.  It is clear that the fresh CNO nuclei produced by 
the SN will, for the lifetime of the SNR, be exposed to a flux of 
energetic particles (EPs) very much higher than the average 
interstellar flux, and this must lead to some spallation production of 
light elements.  This process starts at about \tSW~with a very intense 
radiation field and continues with an intensity decreasing roughly as 
$R^{-3}\propto t^{-6/5}$ (where $R$ is the radius of the SNR) until 
the remnant dies.

The second process is that some of the CNO nuclei from the ejecta are 
accelerated, either by the reverse shock in its brief powerful phase 
at $t\approx\tSW$ or by some of this material managing to get ahead of 
the forward shock.  This later possibility is not impossible, but 
seems unlikely to be as important as acceleration by the reverse 
shock.  Calculations of the Raleigh-Taylor instability of the contact 
discontinuity do suggest that some fast-moving blobs of ejecta can 
punch through the forward shock at about \tSW (Jun \& Norman 1996), 
and in addition Ramaty and coworkers have suggested that fast moving 
dust grains could condense in the ejecta at $t<\tSW$ and then 
penetrate through into the region ahead of the main shock.  In all 
these pictures acceleration of CNO nuclei takes place only at about 
\tSW~and the energy deposited in these accelerated particles is 
certainly less than the explosion energy \ESN, although it might 
optimistically reach some significant fraction of that value (say $\la 
10\%$).  Crucially the accelerated
CNO nuclei are then confined to the interior of the SNR and will thus 
be adiabatically cooled on a rather rapid time-scale, initially of 
order \tSW.

\subsection{Evaluation of the first process (forward shock)}

From the above arguments, it is clear that SNe do induce some Be 
production.  Now the question is~: how much?  Let us first consider 
the irradiation of the ejecta by particles (H and He nuclei) 
accelerated at the forward shock during the Sedov-like phase -- 
process~1.  We have already indicated that detailed studies of 
acceleration in SNRs show that the fraction of the explosion energy 
given to the EPs is roughly constant during the Sedov-like phase and 
of order 0.1 to 0.5 or so.  Let $\theta_{1}$ be that fraction.  Since 
the EPs are distributed more or less uniformly throughout the interior 
of the remnant, the energy density can be estimated as
\begin{equation}
	\EdCR \approx \frac{3\theta_{1}\ESN}{4\pi R^{3}}
	\label{EdCR}
\end{equation}
where $R$ is the radius of the remnant and \ESN~is the explosion 
energy.

To derive a spallation rate from this we need to assume some form for 
the spectrum of the accelerated particles.  Shock acceleration 
suggests that the distribution function should be close to the 
test-particle form $f(p)\propto p^{-4}$ and extend from an injection 
momentum close to `thermal' values to a cut-off momentum at about 
$p_{\mathrm{max}} = 10^{5}\,{\rm GeV}/c$.  The spallation rate per 
target CNO atom to produce a Be atom is then obtained by integrating 
the cross sections
\begin{equation}
	\nu_{\rm spall} = \int_{\pth}^{p_{\mathrm{max}}}
	\sigma v f(p) \d p,
\end{equation}
with the normalisation $\int E(p)f(p)4\pi p^{2}\d p = 
\EdCR$.  Looking at graphs of the spallation cross-sections for Be (as 
given, e.g., in Ramaty et al.  1997), it is clear that these 
cross-sections can be well approximated as zero below a threshold at 
about 30--40~MeV/n and a constant value $\sigma_0\simeq 5\times 
10^{-27}\,\mathrm{cm}^{2}$ above it.  One then obtains roughly~:
\begin{equation}
	\nu_{\rm spall} \approx \frac{\sigma_{0}}{mc} \EdCR
	\frac{1-\ln(\pth/mc)}{1 + \ln(p_{\rm max}/mc)}
	\simeq 0.2 \sigma_0 c \frac{\EdCR}{m c^2},
\end{equation}
where $\pth \simeq mc/5$ is the momentum corresponding to the 
spallation threshold and $m$ refers to the proton mass.  Fortunately, 
for this form of the spectrum the upper cut-off and the spallation 
threshold only enter logarithmically.  A softer spectrum would lead to 
higher spallation yields and a stronger dependence on the spallation 
threshold.

Using Eq.~(\ref{EdCR}) and the adiabatic expansion law for the forward 
shock radius, $R = R_{\mathrm{SW}}(t/\tSW)^{2/5}$, we can now 
estimate the total fraction of the CNO nuclei which will be converted 
to Be during the Sedov-like phase as~:
\begin{eqnarray}
	\phi_{1} &=& \int_{\tSW}^{\tend} 0.2 \sigma_0 c 
	\frac{\theta_{1}\ESN}{mc^{2}} \frac{3}{4\pi R^3} \d t\\
	&=& 0.2 \sigma_0 c \frac{\theta_{1}\ESN}{mc^{2}} \frac{3}{4\pi
	\RSW^3} \int_{\tSW}^{\tend} \left(\frac{t}{\tSW}\right)^{-6/5} dt,
	\label{phi1Interm}
\end{eqnarray}
or
\begin{equation}
	\phi_{1} = \sigma_0 c \frac{\theta_{1}\ESN}{mc^{2}}
	\frac{\rho_{0}}{\Mej} \tSW
	\left[1 - \left(\frac{\tSW}{\tend}\right)^{1/5}\right]
	\label{phi1}
\end{equation}
where as usual $\rho_{0}$ denotes the density of the ambient medium 
into which the SNR is expanding and \Mej~is the total mass of the SNR 
ejecta.  We now recall that the sweep-up time is given in terms of the 
SN parameters and the ambient number density, $n_{0} \approx 
\rho_{0}/m$, as
\begin{equation}
	\tSW = \frac{n_{0}^{-1/3}}{v_{\mathrm{ej}}}
	\left(\frac{3}{4\pi}\frac{M_{\mathrm{ej}}}{m}\right)^{1/3},
\end{equation}
where $v_{\mathrm{ej}}\approx (2\ESN/\Mej)^{1/2}$ is the velocity of 
the ejecta, or numerically~:
\begin{equation}
	\tSW = (1.4\,10^{3}\,\mathrm{yr})
	\left(\frac{M_{\mathrm{ej}}}{10\Msun}\right)^{\hspace{-2pt}\frac{5}{6}}
	\hspace{-4pt}
	\left(\frac{E_{\mathrm{SN}}}{10^{51}\mathrm{erg}}\right)^{\hspace{-2pt}-\frac{1}{2}}
	\hspace{-4pt}
	\left(\frac{n_{0}}{1\mathrm{cm}^{-3}}\right)^{\hspace{-2pt}-\frac{1}{3}}.
	\label{SweepUpTime}
\end{equation}

Replacing in Eq.~(\ref{SweepUpTime}) and using canonical values of $\ESN = 
10^{51}\,\rm erg$ and $\Mej = 10\,\Msun$, we finally get~:

\begin{equation}
	\phi_{1} \simeq 4\times10^{-10}\,\theta_{1} 
	\left(\frac{n_{0}}{1\,\mathrm{cm}^{-3}}\right)^{2/3} \left[1 - 
	\left(\frac{\tSW}{\tend}\right)^{1/5}\right].
	\label{phi1Num}
\end{equation}

Clearly this falls short of the value of order $10^{-8}$ required to 
explain the observations, even for values of $\theta_{1}$ as high as 
$0.5$.  It might seem from Eq.~(\ref{phi1Num}) that very high ambient 
densities could help to make the spallation yields closer to the 
needed value.  This is however not the case.  First, the above 
estimate does not take energy losses into account, while both 
ionisation and adiabatic losses act to lower the genuine production 
rates.  Second, and more significantly, the ratio $\tSW/\tend$ (and 
\emph{a fortiori} its fifth root) becomes very close to 1 in dense 
environments, lowering $\phi_{1}$ quite notably (see 
Fig.~\ref{BeYield}).  In fact, it turns out that there is no 
Sedov-like phase at all in media with densities of order 
$10^{4}\,\mathrm{cm}^{-3}$, the physical reason being that the 
radiative losses then act on a much shorter time-scale, eventually 
shorter than the sweep-up time.

\subsection{Evaluation of the second process (reverse shock)}
\label{SecondProcess}

Let us now turn to the second process, namely the spallation of 
energetic CNO nuclei accelerated at the reverse shock from the SN 
ejecta and interacting within the SNR with swept-up ambient material.  
We have argued above that this reverse shock acceleration is only 
plausible at times around \tSW~and certainly the amount of energy 
transferred to CNO nuclei cannot be more than a fraction of \ESN. Let 
$\theta_2$ be the fraction of the explosion energy that goes into 
accelerating the ejecta at or around \tSW, and $\theta_{\mathrm{CNO}}$ 
the fraction of that energy that is indeed transferred to CNO nuclei.  
These particles are then confined to the interior of the remnant where 
they undergo spallation reactions as well as adiabatic losses.  Let us 
again assume that the spectrum is of the form $f(p)\propto p^{-4}$.  
Then the production rate of Be atoms per unit volume is approximately
\begin{equation}
0.2 {n_{0} \sigma_0\over m c} {\EdCR\over 14}
\label{ProdRate2}
\end{equation}
where \EdCR~now refers to the accelerated CNO nuclei, the factor~14 
comes from the mean number of nucleons per CNO nucleus and the factor 
0.2, as before, from the $f(p)\propto p^{-4}$ spectral shape (assuming 
the same upper cut-off position, but this only enters 
logarithmically).  Integrating over the remnant volume, we obtain the 
spallation rate at \tSW~:
\begin{equation}
	\frac{\d\mathcal{N}_{\mathrm{Be}}}{\d t} \approx 0.2 \sigma_{0}c
	\frac{\theta_{\mathrm{CNO}}\theta_{2}E_{\mathrm{SN}}}{14 m c^{2}}.
	\label{ProdRate2bis}
\end{equation}

Now the adiabatic losses need to be evaluated rather carefully.  It is 
generally argued that they act so that the momentum of the particles 
scales as the inverse of the linear dimensions of the volume occupied.  
Accordingly, in the expanding spherical SNR the EPs should lose 
momentum at a rate $\dot{p}/p = - \dot{R}/R$, reminiscent, 
incidentally, of the way photons behave in the expanding universe.  In 
our case, however, the situation is complicated by the fact that the 
EPs do not push directly against the `walls' limiting the volume of 
confinement, which move at the expansion velocity, $V = \dot{R}$, but 
are reflected off the diffusion barrier consisting of magnetic waves 
and turbulence at rest with respect to the downstream flow, and thus 
expanding at velocity $\frac{3}{4}\dot{R}$.

To see how this influences the actual adiabatic loss rate, it is safer 
to go back to basic physical laws.  Adiabatic losses must arise 
because the EPs are more or less isotropised within the SNR and 
therefore participate to the pressure.  Now this pressure, $P$, works 
positively while the remnant expands, implying an energy loss rate 
equal to the power contributed, given by~:
\begin{equation}
	\frac{\d U}{\d t} = - \int\hspace{-4pt}\int_{\mathcal{S}}\vec{F}\cdot\vec{v}
	= - \int\hspace{-4pt}\int_{\mathcal{S}}P\d S\times\frac{3}{4}\dot{R}
	= - 3\pi R^{2}\dot{R}P,
	\label{KinEnTh}
\end{equation}
where $U = \frac{4}{3}\pi R^{3}\epsilon$ is the total kinetic energy 
of the particles.  Considering that $P = \frac{2}{3}\epsilon$ in the 
non-relativistic limit (NR) and $P = \frac{1}{3}\epsilon$ in the 
ultra-relativistic limit (UR), Eq.~(\ref{KinEnTh}) can be re-writen 
as~:
\begin{equation}
\begin{split}
	\frac{\d\epsilon}{\d t} = -\frac{9}{4}P\frac{\dot{R}}{R}
	&= - \frac{3}{2}\epsilon\frac{\dot{R}}{R}\quad(\mathrm{NR})\\
	&= - \frac{3}{4}\epsilon\frac{\dot{R}}{R}\quad(\mathrm{UR}).
\end{split}
\end{equation}

Finally, dividing both sides by the space density of the EPs and 
noting that $E = p^{2}/2m$ in the NR limit, and $E = pc$ in the UR 
limit, we obtain the momentum loss rate for individual particles, 
valid in any velocity range~:

\begin{equation}
	\frac{\dot{p}}{p} = -\frac{3}{4}\frac{\dot{R}}{R}.
	\label{AdiabLossRate}
\end{equation}

From this one deduces that at the time when the remnant has expanded 
to radius $R$, only those particles whose {\em initial} momenta at 
\tSW~were more than $(R/\RSW)^{3/4}\pth$ are still above the 
spallation threshold.  For a $p^{-4}$ distribution function the 
integral number spectrum decreases as $p^{-1}$ and thus the number of 
accelerated nuclei still capable of spallation reactions decreases as 
$R^{-3/4}\propto t^{-3/10}$.  For a softer accelerated spectrum the 
effect would be even stronger because there are proportionally fewer 
particles at high initial momenta.

This being established, we can integrate Eq.~(\ref{ProdRate2bis}) over 
time, to obtain the total production of Be atoms~:
\begin{equation}
	\mathcal{N}_{\mathrm{Be}} = 0.2 n \sigma_{0} c
	\frac{\theta_{\mathrm{CNO}}\theta_{2}\ESN}{14 mc^{2}}
	\int_{\tSW}^{\tend} \left(\frac{t}{\tSW}\right)^{-3/10}\,\d t
	\label{phi2Interm}
\end{equation}
that is~:
\begin{equation}
	\mathcal{N}_{\mathrm{Be}} = \frac{2}{7} n \sigma_{0} c
	\frac{\theta_{\mathrm{CNO}}\theta_{2}\ESN}{14 mc^{2}} \tSW
	\left[\left(\frac{\tend}{\tSW}\right)^{7/10} - 1\right].
\end{equation}

Dividing by the total number of CNO nuclei in the ejecta, 
$N_{\mathrm{ej,CNO}} = (\theta_{\mathrm{CNO}}/14) N_{\mathrm{ej,tot}} 
\simeq (\theta_{\mathrm{CNO}} M_{\mathrm{ej}}/14 m)$, we get the final 
result~:

\begin{equation}
	\phi_{2} = \sigma_{0}c \frac{\theta_{2}E_{\mathrm{SN}}}{m 
	c^{2}}\frac{\rho_{0}}{M_{\mathrm{ej}}}\tSW
	\frac{2}{7}\left(\frac{\tend}{\tSW}\right)^{7/10}
	\left[1 - \left(\frac{\tSW}{\tend}\right)^{7/10}\right].
	\label{phi2}
\end{equation}
Note that we assumed that the mass fraction of CNO in the ejecta is 
the same as the energy fraction of CNO in the EPs (which was the 
original meaning of $\theta_{\mathrm{CNO}}$).  Considering that all 
nuclear species have the same spectrum in MeV/n, and thus a total 
energy proportional to their mass number, this simply means that the 
acceleration process is not chemically selective, in the sense that 
the composition of the EPs is just the same as that of the material 
passing through the shock.

Numerically, again with $\ESN=10^{51}\,\rm erg$ and $\Mej = 
10\,\Msun$, we finally obtain~:

\begin{equation}
\begin{split}
	\phi_2 \simeq 1\times 10^{-10}&\theta_{2}
	\left(n_0\over 1\,\rm cm^{-3}\right)^{2/3}\\
	&\times\left(\frac{\tend}{\tSW}\right)^{7/10}
	\left[1 - \left(\frac{\tSW}{\tend}\right)^{7/10}\right].
	\label{phi2Num}
\end{split}
\end{equation}

\subsection{Relative contribution of the two processes}
\label{RelativeContribution}

It is worth emphasizing the similarity between expressions 
(\ref{phi1}) and (\ref{phi2}) that we obtained for the spallation 
rates per CNO nuclei by the two processes considered here.  This 
formal analogy allows us to write down their relative contributions 
straightforwardly~:
\begin{equation}
	\frac{\phi_{2}}{\phi_{1}} = \frac{\theta_{2}}{\theta_{1}}
	\times \frac{2}{7}\left(\tend/\tSW\right)^{7/10}
	\frac{\left[1 - \left(\tSW/\tend\right)^{7/10}\right]}
	     {\left[1 - \left(\tSW/\tend\right)^{1/5}\right]}.
	\label{phi2/phi1}
\end{equation}

As is often the case, this similarity is not fortuitous and has a 
physical meaning.  The two processes may indeed be regarded as `dual' 
processes, the first consisting of the irradiation of the SN ejecta by 
the ambient medium, and the second of the ambient medium by the SN 
ejecta.  The `symmetry' is only broken by the dynamical aspect of the 
processes.  First, of course, the energy imparted to the EPs in both 
cases needs not be the same, for it depends on the acceleration 
efficiency as well as the total energy of the shock involved (forward 
or reverse).  This is expressed by the expected ratio 
$\theta_{2}/\theta_{1}$.  And secondly, in the first process one has 
to fight against the dilution of the ejecta -- integration of 
$(t/\tSW)^{-6/5}$, see Eq.~(\ref{phi1Interm}) -- while in the second 
process one fights against the adiabatic losses -- integration of 
$(t/\tSW)^{-3/10}$, see Eq.~(\ref{phi2Interm}).  This is expressed by 
the last factor in Eq.~(\ref{phi2/phi1}).

Clearly the latter decrease of the production rates is the least 
dramatic, and the reverse shock process must dominate the LiBeB 
production in supernova remnants.  However, this conclusion still 
depends on the genuine efficiency of reverse shock acceleration, and 
once the relative acceleration efficiency $\theta_{2}/\theta_{1}$ is 
given, the weight of the first process relative to the second still 
depends on the total duration of the Sedov-like phase, appearing 
numerically in Eq.~(\ref{phi2/phi1}) through the ratio \tend/\tSW, 
which in turn depends on the ambient density, $n_{0}$.  The expression 
of \tSW~as a function of the parameters has been given in 
Eq.~(\ref{SweepUpTime}), so we are left with the evaluation of the time, 
\tend, when the magnetic turbulence collapses and the EPs leave the 
SNR, putting an end to Be production.  We argued above that 
\tend~should correspond to the end the Sedov-like phase, when the 
shock induced by the SN explosion becomes radiative, that is when the 
cooling time of the post-shock gas becomes of the same order as the 
dynamical time.

In principle, the cooling rate can be derived from the so-called 
cooling function, $\Lambda(T) (\mathrm{erg~cm}^{3}\mathrm{s}^{-1}$), 
which depends on the physical properties of the post-shock material, 
notably on its temperature, $T$, and metallicity, $Z$~:
\begin{equation}
	\tau_{\mathrm{cool}}\approx\frac{\frac{3}{2}k_{\mathrm{B}}T}{n\Lambda(T)},
	\label{tauCool}
\end{equation}
where $n$ is the post-shock density, equal to $4n_{0}$ if the 
compression ratio is that of an ideal strong shock (nonlinear effects 
probably act to increase the compression ratio to values larger 
than~4).  As for the dynamical time, we simply write
\begin{equation}
	\tau_{\mathrm{dyn}} \approx \frac{\dot{R}}{R} \approx \frac{5}{2}\,t.
	\label{tauDyn}
\end{equation}

To obtain \tend, we then need to solve the following equation in the 
variable $t$, obtained by equating $\tau_{\mathrm{cool}}$ and 
$\tau_{\mathrm{dyn}}$ given above~:
\begin{equation}
	t \approx \frac{3k_{\mathrm{B}}T}{20n_{0}\Lambda(T)},
	\label{tEndEq}
\end{equation}
where it should be clear that the right hand side also depends on time 
through the temperature, $T$, and thus indirectly through the cooling 
function too.  In the non-radiative SNR expansion phase, the function 
$T(t)$ is obtained directly from the hydrodynamical jump conditions at 
the shock discontinuity~:
\begin{equation}
	T \approx \frac{3m}{8k_{\mathrm{B}}}V^{2},
\end{equation}
or numerically~:
\begin{equation}
	T \approx (2\times 10^{5}\,\mathrm{K})
	\left(\frac{E_{\mathrm{SN}}}{10^{51}\mathrm{erg}}\right)^{\hspace{-2pt}\frac{2}{5}}
	\hspace{-4pt}
	\left(\frac{n_{0}}{1\mathrm{cm}^{-3}}\right)^{\hspace{-2pt}-\frac{2}{5}}
	\hspace{-4pt}
	\left(\frac{t}{10^{5}yr}\right)^{\hspace{-2pt}-\frac{6}{5}}.
	\label{Temperature}
\end{equation}

%%%%%%%%%%%%%%
\begin{figure}
\resizebox{\hsize}{!}{\includegraphics{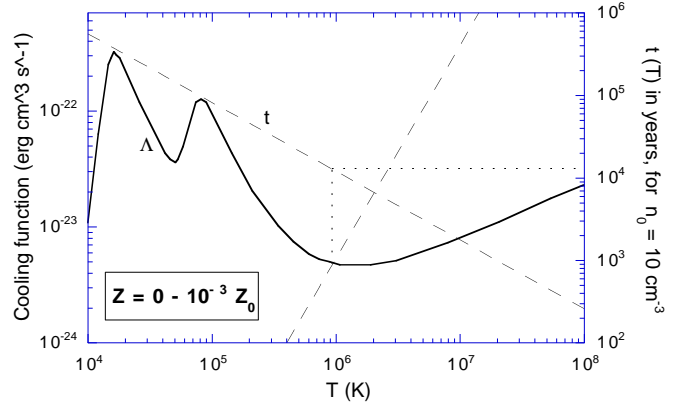}} 
\caption{Cooling function (bold line) as a function of the temperature 
for a medium with metallicity lower than $\sim 10^{-2}Z_{\odot}$.  The 
dashed lines illustrate the graphical determination of \tend, for an 
ambient density $n_{0} = 10\,\mathrm{cm}^{-3}$ (see text).}
\label{CoolingFunction}
\end{figure}
%%%%%%%%%%%%%%

To solve Eq.~(\ref{tEndEq}), we still need to know the cooling 
function $\Lambda(T)$.  In the range of temperatures corresponding to 
the end of the Sedov-like phase, $10^{5}\,\mathrm{K}\la T\la 
10^{7}\,\mathrm{K}$, it happens to depend significantly on 
metallicity, with differences up to two orders of magnitude for 
metallicities from $Z = 0$ to $Z = 2 Z_{\odot}$ (B\"oh\-rin\-ger \& 
Hens\-ler 1989).  Because we focus on Be production in the early 
Galaxy, we adopt the cooling function corresponding to zero 
metallicity, represented in Fig.~\ref{CoolingFunction} (adapted from 
B\"oh\-rin\-ger \& Hens\-ler 1989), which holds for values of $Z$ up 
to $\sim 10^{-2}Z_{\odot}$.

For high enough ambient densities, the shock will become radiative 
early in the SNR evolution, when the temperature is still very high, 
say above $T\ga 2\,10^{6}$~K. In this case, the cooling function is 
dominated by Bremsstrahlung emission and can be written analytically 
as~:
\begin{equation}
	\Lambda_{\mathrm{Br}}(T) \approx 
	(2.4\,10^{-23}\,\mathrm{erg~cm}^{3}\mathrm{s}^{-1})
	\left(\frac{T}{10^{8}\,\mathrm{K}}\right)^{1/2}.
	\label{LambdaBr}
\end{equation}
Substituting from (\ref{Temperature}) and (\ref{LambdaBr}) in 
Eq.~(\ref{tEndEq}) and solving for $t$, we find~:
\begin{equation}
	\tend = (1.1\,10^{5}\,\mathrm{yr})
	\left(\frac{E_{\mathrm{SN}}}{10^{51}\mathrm{erg}}\right)^{1/8}
	\left(\frac{n_{0}}{1\mathrm{cm}^{-3}}\right)^{-3/4}.
	\label{tEnd}
\end{equation}

\begin{figure}
\resizebox{\hsize}{!}{\includegraphics{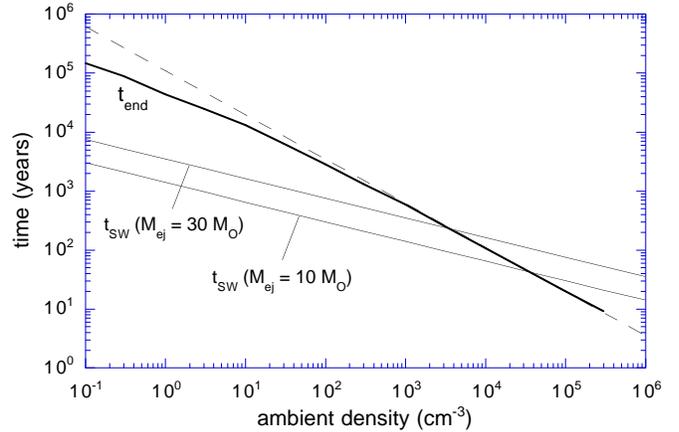}} 
\caption{Comparison of \tend~and \tSW~as a function of the ambient 
density, $n_{0}$, for two different values of the mass ejected by the 
supernova (10 and 30~\Msun). The dashed line shows the asymptotic 
analytic estimate of Eq.~(\ref{tEnd}).}
\label{TimeScales}
\end{figure}

To check the consistency of our assumption $\Lambda \approx 
\Lambda_{\mathrm{Br}}$ (i.e.  $T\ga 2\,10^{6}$~K), let us now report 
Eq.~(\ref{tEnd}) in (\ref{Temperature}) and write down the temperature 
$T_{\mathrm{End}}$ at the end of the Sedov-like phase~:
\begin{equation}
	T_{\mathrm{End}} \approx (2\times 10^{5}\,\mathrm{K})
	\left(\frac{E_{\mathrm{SN}}}{10^{51}\mathrm{erg}}\right)^{1/4}
	\left(\frac{n_{0}}{1\mathrm{cm}^{-3}}\right)^{1/2},
\end{equation}
which means that the above analytical treatment is valid only for 
ambient densities greater than about $100\,\mathrm{cm}^{-3}$.  For 
lower densities, we must solve Eq.~(\ref{tEndEq}) graphically.  First, 
we invert Eq.~(\ref{Temperature}) to express $t$ as a function of 
temperature, then we plot the function $f(T)\equiv 
3k_{\mathrm{B}}T/(20n_{0}t)$ on the same graph as $\Lambda$ (see 
Fig.~\ref{CoolingFunction} for an example), find the value of $T$ at 
intersection, and finally convert this value into the sought time 
\tend~making use again of Eq.~(\ref{Temperature}).  The results, 
showing \tend~as a function of the ambient density, are shown in 
Fig.~\ref{TimeScales}.

%%%%%%%%%%%%%%
%%%%%%%%%%%%%%

\begin{figure}
\resizebox{\hsize}{!}{\includegraphics{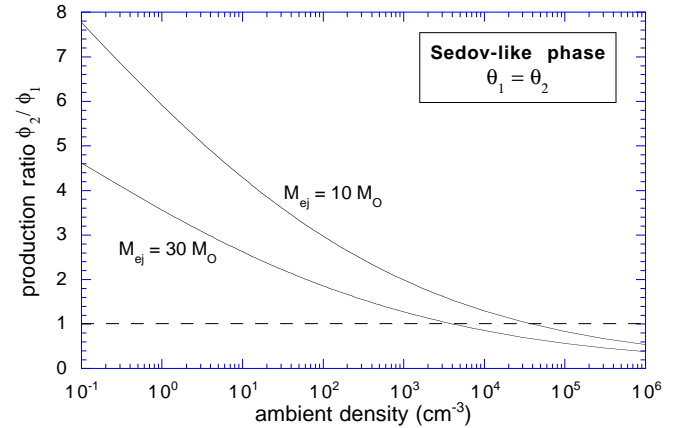}} 
\caption{Comparison of the Be (and B) production efficiency through 
the forward and reverse shock acceleration processes, for two values 
of the ejected mass (10~and 30~\Msun).  The ratio $\phi_{2}/\phi_{1}$ 
(see text) is plotted as a function of the ambient density, assuming 
that both processes impart the same total energy to the EPs 
($\theta_{1} = \theta_{2}$).}
\label{Phi2/Phi1}
\end{figure}

We now have all the ingredients to plot the efficiency ratio of the 
two processes calculated above.  Figure~\ref{Phi2/Phi1} shows the 
ratio $\phi_{2}/\phi_{1}$ given in Eq.~(\ref{phi2/phi1}) as a function 
of the ambient density, assuming that $\theta_{1} = \theta_{2}$.  Two 
different values of the ejected mass have been used, corresponding to 
different progenitor masses ($\sim 10-40\,\Msun$).  It can be seen 
that low densities are more favourable to the reverse shock 
acceleration process.  This is due to $\tend/\tSW$ being larger, 
implying a larger dilution of the ejecta (process~1 less efficient) 
and smaller adiabatic losses, which indeed decrease as $t^{-1}$ 
(process~2 more efficient).  The part of the plot corresponding to 
$\phi_{2}/\phi_{1}\le 1$ is not physical, because it requires 
$\tend\le\tSW$, which simply means that the Sedov-like phase no longer 
exists and the whole calculation becomes groundless.  Note however 
that in Fig.~\ref{Phi2/Phi1} the energy imparted to the EPs has been 
assumed equal for both processes, which is most certainly not the 
case.  Actually, if $\theta_{2}/\theta_{1} = 0.1$ (e.g.  $\theta_{1} = 
10\%$ and $\theta_{1} = 1\%$), then process~1 is found to dominate Be 
production during the Sedov-like phase, regardless of the ambient 
density.

%%%%%%%%%%%%%%
%%%%%%%%%%%%%%

\section{Spallation reactions after the Sedov-like phase}

At the end of the Sedov-like phase, the EPs are no longer confined and 
leave the SNR to diffuse across the Galaxy.  At the stage of chemical 
evolution we are considering here, there are no or few metals in the 
interstellar medium (ISM), so that energetic protons and $\alpha$ 
particles accelerated at the forward shock will not produce any 
significant amount of Be after \tend~(although Li production will 
still be going on through $\alpha + \alpha$ reactions).  In the case 
of the second process, however, the EPs contain CNO nuclei which just 
cannot avoid being spalled while interacting with the ambient H and He 
nuclei at rest in the Galaxy.  This may be regarded as a third process 
for Be production, which lasts until either the EPs are slowed down by 
Coulombian interactions to subnuclear energies (i.e.  below the 
spallation thresholds) or they simply diffuse out of the Galaxy.  
Since the confinement time of cosmic rays in the early Galaxy is 
virtually unknown, we shall assume here that the Galaxy acts as a 
thick target for the EPs leaving the SNR, an assumption which actually 
provides us with an upper limit on the spallation yields.

Unlike the first two processes evaluated above, this third process is 
essentially independent of dynamics.  Thus, time-dependent 
calculations are no longer needed and, from this stage on, the 
calculations made by Ramaty et~al.  (1997) or any steady-state 
calculation is perfectly valid.  In particular, the ambient density 
has no influence on light element production, since a greater number 
of reactions per second, as would result from a greater density, 
implies an equal increase of both the spallation rates and the energy 
loss rate.  Once integrated over time, both effects cancel out 
exactly, and in fact, given the energy spectrum of the EPs, the 
efficiency of Be production (and Li, and B), expressed as the number 
of nuclei produced per erg injected in the form of EPs, depends only 
on their chemical composition.

\begin{figure}
\resizebox{\hsize}{!}{\includegraphics{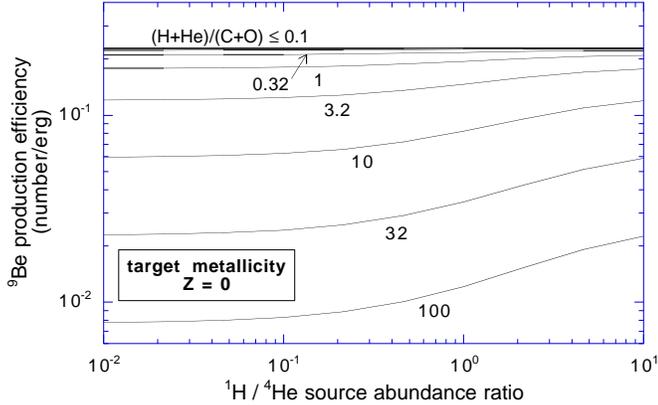}} 
\caption{Production efficiency of Be, as a function of the EP 
composition (all abundances are by number).  The ordinate is the 
number of Be nuclei produced by spallation reactions per erg injected 
in the form of EPs.  A thick target has been assumed, with zero 
metallicity.  Carbon and Oxygen abundances were set equal in the EP 
composition.}
\label{Be/erg}
\end{figure}

Results are shown in Fig.~\ref{Be/erg} for different values of the 
source abondance ratios, $\mathrm{H}/\mathrm{He}$ and 
$(\mathrm{H}+\mathrm{He})/(\mathrm{C}+\mathrm{O})$, allowing one to 
derive the spallation efficiency for any composition.  Two-steps 
processes (such as $^{12}\mathrm{C} + ^{1}\mathrm{H} \longrightarrow 
^{10}\mathrm{B}$ followed by $^{10}\mathrm{B} + ^{1}\mathrm{H} 
\longrightarrow ^{9}\mathrm{Be}$) have been taken into account.  Test 
runs show good agreement with the results of Ramaty et~al.  (1997).

%%%%%%%%%%%%%%
%%%%%%%%%%%%%%

As can be seen on Fig.~\ref{Be/erg}, pure Carbon and Oxygen have a 
production efficiency of about $0.22\,\mathrm{nuclei}/\mathrm{erg}$, 
while this efficiency decreases by at least a factor of~10 for 
compositions with hundred times more H and He than metals (or about 
ten times more by mass).  According to models of explosions for SN 
with low metallicity progenitors, the average 
$(\mathrm{H}+\mathrm{He})/(\mathrm{C}+\mathrm{O})$ ratio among the EPs 
should indeed be expected to be $\ga 200$, unless selective 
acceleration occurs to enhance the abundance of the metals.  As a 
consequence, efficiencies greater than $\sim 
10^{-2}\,\mathrm{nuclei}/\mathrm{erg}$ should not be expected, so that 
a production of $\sim 4\,10^{48}$~atoms of Be requires an energy of 
$\sim 4\,10^{50}$~erg to be imparted to the EPs.  This seems very 
unlikely considering that the total energy available in the reverse 
shock (the source of the EPs) should be of order one tenth of the SN 
explosion energy, not to mention the acceleration efficiency.  
Moreover, a significant fraction of the energy originally imparted to 
the EPs has been lost during the Sedov-like phase of the SNR evolution 
through adiabatic losses.

%%%%%%%%%%%%%%
\begin{figure}
\resizebox{\hsize}{!}{\includegraphics{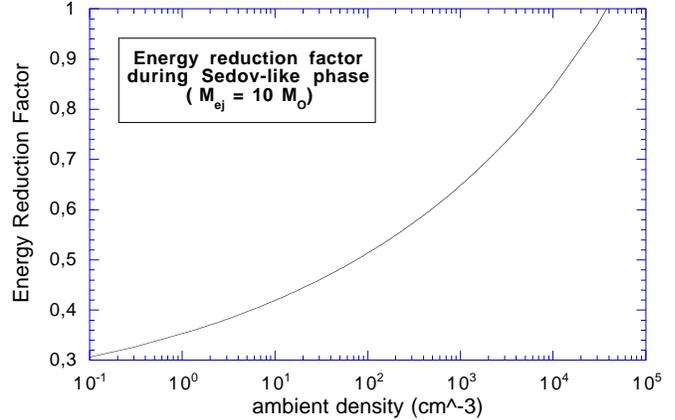}} 
\caption{Fraction of the energy imparted to the EPs at time \tSW~which 
is still available at \tend, after the Sedov-like phase, as a function 
of othe ambient density.}
\label{EnergyReduction}
\end{figure}
%%%%%%%%%%%%%%

To evaluate the `surviving' fraction of energy, it sufficies to go 
back to Eq.~(\ref{AdiabLossRate}), which indicates that when the 
radius of the shock is multiplied by a factor $\eta$, the momentum $p$ 
of all the particles is multiplied by a factor $\eta^{-3/4}$.  It is 
worthwhile noting that, because of their specific momentum dependence, 
adiabatic losses do not modify the shape of the EP energy spectrum.  
In our case, $f(p)\propto p^{-4}$, so that when all momenta $p$ are 
divided by a factor $\zeta$, the distribution function $f(p)$ is 
divided by the same factor $\zeta$.  To see that, the easiest way is 
to work out the number of particules between momenta $p$ and $p + \d 
p$ after the momentum scaling.  This number writes $\d N^{\prime} = 
f^{\prime}(p)4\pi p^{2}\d p$, where $f^{\prime}(p)$ is the new 
distribution function.  Now $\d N^{\prime}$ must be equal to the 
number of particles that had momentum between $\zeta p$ and $\zeta(p + 
\d p)$, which is, by definition, $\d N = f(\zeta p)4\pi(\zeta 
p)^{2}\zeta\d p$.  Equating $\d N$ and $\d N^{\prime}$ yields the 
result $f^{\prime}(p) = \zeta^{-1}f(p)$.

Putting all pieces together, we find that when the shock radius $R$ is 
multiplied by a factor $\eta$, the distribution function and, thus, 
the total energy of the EPs are multiplied by $\eta^{-3/4}$.  Now 
considering that $R$ increases as $t^{2/5}$ during the Sedov-like 
phase, we find that the total energy of the EPs decreases as 
$t^{-3/10}$.  Note that this is nothing but an other way to work out 
the decrease of the spallation rates for our second process during the 
Sedov-like phase (cf.  Sect.~\ref{SecondProcess}).  Finally, we find 
that a fraction $(\tend/\tSW)^{-3/10}$ of the initial energy imparted 
to the EPs is still available for spallation at the end of the 
Sedov-like phase.  This factor is plotted on 
Fig.~\ref{EnergyReduction}, as a function of the ambient density.  It 
can be seen that for $n_{0} = 1\,\mathrm{cm}^{-3}$, the energy 
available to power our third process of light element production has 
been reduced by adiabatic losses to not more than one third of its 
initial value, and less than one half for densities up to 
$100\,\mathrm{cm}^{-3}$.  Clearly, high densities are favoured 
(energetically) because they tend to shorten the Sedov-like phase, and 
therefore merely avoid the adiabatic losses.

\section{Discussion}

Since light element production in the interstellar medium obviously 
requires a lot of energy in the form of supernuclear particles (i.e.  
with energies above the nuclear thresholds) as well as metals 
(especially C and O), it is quite natural to consider SNe as possible 
sources of the LiBeB observed in halo stars.  We have analysed in 
detail the spallation nucleosynthesis induced by a SN explosion on the 
basis of known physics and theoretical results relating to particle 
shock acceleration.  Two major processes can be identified, depending 
on whether the ISM or the ejecta are accelerated, respectively at the 
forward and reverse shocks.  In the first case, the EPs consist mostly 
of protons and alpha particles and must therefore interact with C and 
O nuclei, which are much more numerous within the SNR than in the 
surrounding medium (especially at early stages of Galactic evolution).  
The process will thus last as long as the EPs stay confined in the 
SNR, i.e.  approximately during the Sedov-like phase, but not more.  
In the second case, freshly synthesized CNO nuclei are accelerated, 
and Be production occurs through interaction with ambient H and He 
nuclei. The process is then divided into two, one stretching over the 
Sedov-like phase, with the particles suffering adiabatic losses, and 
the other one occuring outside the remnant, with only Coulombian 
losses playing a role.

%%%%%%%%%%%%%%
\begin{figure}
\resizebox{\hsize}{!}{\includegraphics{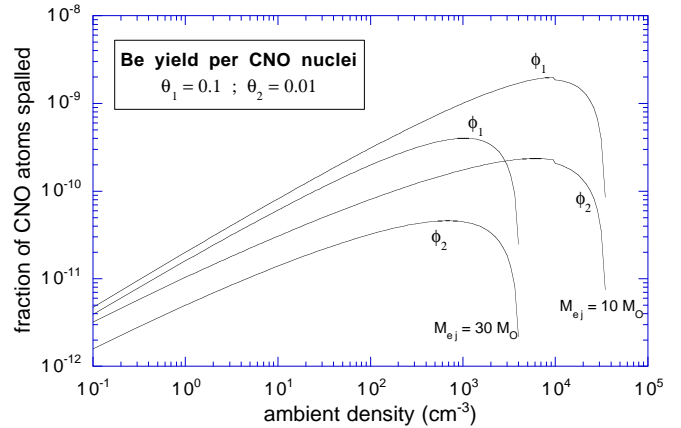}} 
\caption{Spallation efficiency of CNO during the Sedov-like phase, as 
a function of the ambient density.  The fraction of freshly 
synthesized CNO nuclei being spalled to Be by processes~1 ($\phi_{1}$) 
and~2 ($\phi_{2}$) is obtained from Eqs.~(\ref{phi1Num}) 
and~(\ref{phi2Num}) and the values of \tend~derived in 
Sect.~\ref{RelativeContribution}, for two values of the ejected mass 
(10~and 30~\Msun).}
\label{BeYield}
\end{figure}
%%%%%%%%%%%%%%

We have calculated the total Be production in these three processes, 
taking the dynamics of the SNR evolution into account (dilution of the 
ejecta by metal-poor material and adiabatic losses).  The results are 
shown in Fig.~\ref{BeYield} for processes 1 and~2 (from 
Eqs.~(\ref{phi1Num}) and~(\ref{phi2Num})).  We find that with 
canonical values of $\theta_{1} = 0.1$, $\theta_{2} = 0.01$, 
$M_{\mathrm{ej}} = 10\,\Msun$ and a mean ambient density $n_{0} = 
10\,\mathrm{cm}^{-3}$, the fraction of freshly synthesized CNO nuclei 
spalled into Be in these processes is $\phi_{1} \sim 8\,10^{-11}$ and 
$\phi_{2} \sim 3\,10^{-11}$, respectively, which is very much less 
than the value `required' by the observations, discussed in the 
introduction ($\phi_{\mathrm{obs}}\sim 3\,10^{-8}$).  Even allowing 
for unreasonably high values of the acceleration efficiency, 
$\theta_{1}\sim \theta_{2}\la 1$, the total Be production by processes 
1 and~2 would still be more than one order of magnitude below the 
observed value.

As suggested by Fig.~\ref{BeYield} and our analytical study, higher 
densities improve the situation.  However, even with $n_{0} = 
10^{3}\,\mathrm{cm}^{-3}$ and acceleration efficiencies equal to~1, 
the Be yield is still unsufficient.  Moreover, it should be noted that 
our calculations did not consider Coulombian energy losses (because 
they are negligeable as compared to adiabatic losses for usual 
densities), which become important as the density increases and 
therefore make the Be yield smaller.  Finally, since we are trying to 
account for the mean abundance of Be in halo stars, as compared to Fe, 
we have to evaluate the Be production for an ambient density 
corresponding to the mean density encountered around explosion sites 
in the early Galaxy, which is very unlikely to be as high as 
$10^{3}\,\mathrm{cm}^{-3}$.  It could even be argued that although the 
gas density might have been higher in the past than it is now (hence 
our `canonical value' $n_{0} = 10\,\mathrm{cm}^{-3}$), the actual mean 
density about SN explosion sites could be lower than 
$1\,\mathrm{cm}^{-3}$, because most SNe may explode within superbubble 
interiors, where the density is much less than in the mean ISM.

Thus, our conclusion is that processes 1 and~2 both fail in accounting 
for the Be observed in metal-poor stars in the halo of our Galaxy.  
Concerning the third process, adopting canonical values for the 
parameters again leads to unsufficient Be production, as noted in the 
previous section.  While higher densities improve the situation by 
avoiding the adiabatic energy losses, one should nevertheless expect 
at least half of the EP energy to be lost in this way, for any 
reasonable density.  This means that even if 10\%
of the explosion energy is imparted to EPs accelerated at the reverse 
shock, which is certainly a generous upper limit, the required number 
of $\sim 4\,10^{48}$ nuclei of Be per SN implies a spallation 
efficiency of $\sim 0.1$~nucleus/erg.  Now Fig.~\ref{Be/erg} shows 
that this requires an EP composition in which at least one particle 
out of ten is a CNO nucleus.  In other words, the ejected mass of CNO 
must be of the order of that of H and He together.  None of the SN 
explosion models published so far can reproduce such a requirement, 
and so there is clearly a problem with Be production in the early 
Galaxy.

The results presented here are in fact interesting in many regards.  
First, they show that it is definitely very difficult to account for 
the amount of Be found in halo stars.  Consequently, we feel that the 
main problem to be addressed in this field of research is probably not 
the chemical evolution of Be (and Li and B) in the Galaxy, as given by 
the ensemble of the data points in the abundance vs metallicity 
diagrams (e.g.  whether Be is proportional to Fe or to its square) 
but, to begin with, the position of any of these points.  Are we able 
to describe in some detail one process which could explain the amount 
of Be (relative to Fe) present in any of the stars in which it is 
observed?  The answer, we are afraid, seems to be no at this stage.  
It is however instructive to ask why the processes investigated here 
have failed.  Concerning process~1 (acceleration of ISM, interaction 
with fresh CNO within the SNR), the main reason is that the CNO rich 
ejecta are `too much diluted' by the swept-up material as the SNR 
expands, so that the spallation efficiency is too low (or the 
available energy is too small).  However, it seems rather hard to 
think of any region in the Galaxy where the concentration in CNO is 
higher than inside a SNR during the Sedov-like phase (especially in 
the first stages of chemical evolution)!  So the conclusion that 
process~1 cannot work, even with a 100\% acceleration
efficiency, seems to rule out any other process based on the 
acceleration of the ISM, initially devoided of metals.

The other solution is then of course to accelerate CNO nuclei 
themselves, which provides the maximum possible spallation efficiency, 
independently of the ambient metallicity.  Every energetic CNO will 
lead to the production of as much Be as possible given the spallation 
cross sections and the energy loss rates.  The latter cannot 
physically be smaller than the Coulombian loss rate in a neutral 
medium, and this leads to the efficiency plotted in Fig.~\ref{Be/erg}.  
Unfortunately, a significant amount of the CNO rich ejecta of an 
isolated SN can only be accelerated at the reverse shock at a time 
around the sweep-up time, \tSW. This means that i) the total amount of 
energy available is smaller than the explosion energy (probably of 
order 10\%, i.e.  $\sim 10^{50}$~erg), and ii) the accelerated nuclei
will suffer adiabatic losses during the Sedov-like phase, reducing 
their energy by a factor of 2 or~3. As shown above, this makes 
process~2-3 incapable of producing enough Be, as long as the EPs have 
a composition reflecting that of the SN ejecta.

This suggest that a solution to the problem could be that the reverse 
shock accelerates preferentially CNO nuclei rather than H and He.  For 
example, recent calculations have shown that such a selective 
acceleration arises naturally if the metals are mostly condensed in 
grains (Ellison et~al.  1997).  The proposition by Ramaty et~al.  
(1997) that grains condense in the ejecta before being accelerated 
could then help to increase the abundance of CNO in the EPs.  However, 
we have to keep in mind that any selective process called upon must be 
very efficient indeed, since as we indicated above, the data require 
that the EP composition be as rich as one CNO nuclei out of ten EPs, 
which means that CNO nuclei must be accelerated at least ten times 
more efficiently than H and He.  This would have to be increased by 
another factor of ten if the energy initially imparted to the EPs by 
the acceleration process were only a factor 2 or~3 lower (i.e.  $\sim 
3\,10^{49}$~erg, which is more reasonable from the point of view of 
particle acceleration theory).  Clearly, more work is needed in this 
field before one can safely invoke a solution in terms of selective 
acceleration.

As can be seen, playing with the composition to increase the 
spallation efficiency has its own limits, and in any case, 
Fig.~\ref{Be/erg} gives an unescapable upper limit, obtained with pure 
Carbon and Oxygen (at least for the canonical spectrum considered here 
- other spectra were also investigated, as in Ramaty et al.  (1997), 
leaving the main conclusions unchanged).  This would then suggest that 
another source of energy should be sought.  However, the constraint 
that it should be more energetic than SNe is rather strong.

Another interesting line of investigations could be the study of the 
collective effects of SNe.  Most of the massive stars and SN 
progenitors are believed to be born (and indeed observed, Melnik 
\&~Efremov 1995) in associations, and their joint explosions lead to 
the formation of superbubbles which may provide a very favourable 
environment for particle acceleration (Bykov \&~Fleishman 1992).  
Parizot et~al.  (1998) have proposed that these superbubbles could be 
the source of most of the CNO-rich EPs, and Parizot \&~Knoedlseder 
(1998) further investigated the gamma-ray lines induced by such an 
energetic component.  The most interesting features of a scenario in 
which Be-producing EPs are accelerated in superbubbles is that i) when 
a new SN explodes, the CNO nuclei ejected by the previous SNe are 
accelerated at the \emph{forward shock}, instead of the reverse shock 
in the case of an isolated SN, which implies a greater energy, and ii) 
no significant adiabatic losses occur, because of the dimensions and 
low expansion velocity of the superbubble.  This makes the superbubble 
scenario very appealing, and it will be investigated in detail in a 
forthcoming paper.

However that may be, we should also keep in mind that when we say that 
a process does not produce enough Be, it always means that it does not 
produce enough Be \emph{as compared to Fe}.  Now it could also be that 
SN explosion models actually produce too much Fe.  The point is that 
Be is compared to Fe in the observations, while it has no direct 
physical link with it.  Indeed, Be is not made out of Fe, but of C 
and~O. So to be really conclusive, the studies of spallative 
nucleosynthesis should compare theoretical Be/O yields to the 
corresponding abundance ratio in metal-poor stars.  Unfortunately, the 
data are much more patchy for Be as a function of [O/H] than as a 
function of [Fe/H], especially in very low metallicity stars.  The 
usually assumed proportionality between O and Fe could turn out to be 
only approximate, as recent observational works possibly indicate 
(Israelian et~al.  1998; Boesgaard et~al.  1998; these observations, 
however, still ??  need to be confirmed by an independent method, all 
the more that they come into conflict with several theoretical and 
observational results; cf.  Vangioni-Flam et~al.  1998b).  We shall 
address this question in greater detail in the attending paper 
(Parizot and Drury, 1999, Paper~II).

Finally, we wish to stress that the calculations presented in this 
paper rely on a careful account of the dynamics of the problem.  More 
generally, time-dependent calculations are required to properly 
evaluate the spallation processes in environments where compositions 
and energy densities are evolving.  In particular, as argued in 
Parizot (1998), no variation with density can be obtained with a 
stationary model, since an increase in the density induces an 
equivalent and cancelling increase in the spallation rates and the 
energy loss rates.  By contrast, we have shown that all three of the 
processes considered here are more efficient at higher density -- a 
result which could not have been found otherwise.  Detailed, numerical 
time-dependent calculations will be presented in paper~II, with 
conclusions similar to those demonstrated here.

\begin{acknowledgements}
This work was supported by the TMR programme of the European Union 
under contract FMRX-CT98-0168.  It was initiated during a visit by LD 
to the Service d'Astrophysique, CEA Saclay, whose hospitality is 
gratefully acknowledged.  We wish to thank M. Cass\'e and Elisabeth 
Vangioni-Flam for stimulating discussions of these and related topics.
\end{acknowledgements}

\end{document}